\documentclass[aps,prl,twocolumn,showpacs,superscriptaddress]{revtex4}

\usepackage{graphicx}
\usepackage[english]{babel}
\usepackage{bm}    

\def\smfrac#1#2{{\textstyle\frac{#1}{#2}}}
\def\smhalf{ {\smfrac{1}{2}} }
\def\half{ {\frac{1}{2}} }

\def\reff#1{(\ref{#1})}
\newcommand{\be}{\begin{equation}}
\newcommand{\ee}{\end{equation}}
\newcommand{\<}{\langle}
\renewcommand{\>}{\rangle}

\def\spose#1{\hbox to 0pt{#1\hss}}
\def\ltapprox{\mathrel{\spose{\lower 3pt\hbox{$\mathchar"218$}}
 \raise 2.0pt\hbox{$\mathchar"13C$}}}
\def\gtapprox{\mathrel{\spose{\lower 3pt\hbox{$\mathchar"218$}}
 \raise 2.0pt\hbox{$\mathchar"13E$}}}

\newcommand{\scrc}{{\cal C}}

\newcommand{\scrn}{{\cal N}}
\newcommand{\scro}{{\cal O}}

\newcommand{\scrs}{{\cal S}}

\begin{document}

\title{Critical speeding-up in a local dynamics for the random-cluster model}

\author{Youjin Deng}
\affiliation{Department of Physics, New York University,
      4 Washington Place, New York, NY 10003, USA}
\author{Timothy M.~Garoni}
\affiliation{Department of Physics, New York University,
      4 Washington Place, New York, NY 10003, USA}
\author{Alan D. Sokal}
\affiliation{Department of Physics, New York University,
      4 Washington Place, New York, NY 10003, USA}
\affiliation{Department of Mathematics,
      University College London, London WC1E 6BT, UK}

\date{January 5, 2007}

\begin{abstract}
We study the dynamic critical behavior
of the local bond-update (Sweeny) dynamics for the
Fortuin--Kasteleyn random-cluster model in dimensions $d=2,3$,
by Monte Carlo simulation.
We show that, for a suitable range of $q$ values,
the global observable $\scrs_2$ exhibits ``critical speeding-up'':
it decorrelates well on time scales much less than one sweep,
so that the integrated autocorrelation time tends to zero
as the critical point is approached.
We also show that the dynamic critical exponent $z_{\rm exp}$
is very close (possibly equal) to the rigorous lower bound $\alpha/\nu$,
and quite possibly smaller than the corresponding exponent
for the Chayes--Machta--Swendsen--Wang cluster dynamics.
\end{abstract}

\pacs{05.50.+q, 05.10.Ln, 05.70.Jk, 64.60.Ht}

\keywords{Dynamic critical phenomena, critical slowing-down,
   critical speeding-up, Potts model, Fortuin--Kasteleyn representation,
   random-cluster model, Sweeny algorithm, Monte Carlo.}

\maketitle


Dynamic processes in statistical mechanics typically undergo
{\em critical slowing-down}\/ \cite{Hohenberg_77}:
the autocorrelation (relaxation) time $\tau$ diverges
as the critical point is approached,
most often like $\tau \sim \xi^z$,
where $\xi$ is the spatial correlation length
and $z$ is a dynamic critical exponent.
In this Letter we would like to draw attention to the converse
(and quite unexpected, at least to us)
phenomenon of {\em critical speeding-up}\/:
some observables $\scro$ can exhibit strong decorrelation
on time scales much {\em less}\/ than one sweep,
so that the dominant relaxation modes
equilibrate {\em faster}\/
(in natural time units)
near criticality.
As a consequence,
the integrated autocorrelation time $\tau_{{\rm int},\scro}$
can in some cases tend to {\em zero}\/,
so that the dynamic critical exponent $z_{{\rm int},\scro}$ is {\em negative}\/.
These behaviors also have practical implications
for the efficiency of Monte Carlo simulations
\cite{Binder_79-92,Sokal_Cargese_96}
near the critical point.

More precisely, we shall exhibit these phenomena for the
Fortuin--Kasteleyn (FK) random-cluster model \cite{FK_69+72,Grimmett_06}
with a local bond-update dynamics \cite{Sweeny_83}.
The random-cluster model is a correlated bond-percolation model
that is closely related to the Potts spin model \cite{Potts_52,Wu_82+84}.
As such, it plays a major role in the theory of critical phenomena,
especially in two dimensions where it arises
in recent developments of conformal field theory \cite{DiFrancesco_97}
via its connection with stochastic Loewner evolution (SLE)
\cite{SLE_math,SLE_phys}.

The random-cluster model with parameters $q,v > 0$
is defined on any finite graph $G=(V,E)$ by the partition function
\begin{equation}
   Z \;=\;  \sum_{A \subseteq E} q^{k(A)} v^{|A|}
   \;,
 \label{eq.ZRC}
\end{equation}
where $A$ is the set of ``occupied bonds'',
$|A|$ is the number of occupied bonds,
and $k(A)$ is the number of connected components (``clusters'')
in the graph $(V,A)$.
For $q=1$ this reduces to independent bond percolation \cite{Stauffer_92}
with occupation probability $p = v/(1+v)$;
for integer $q \ge 2$ it provides a graphical representation \cite{FK_69+72}
of the $q$-state ferromagnetic Potts model
with nearest-neighbor coupling $J$, where $v = e^{\beta J} - 1$.
The random-cluster model thus provides an extension of the Potts model
that allows all positive values of $q$, integer or noninteger,
to be studied within a unified framework.

The simplest dynamics for the random-cluster model
is the local bond-update dynamics
first used by Sweeny \cite{Sweeny_83}:
choose a bond $e \in E$ at random, erase its current occupation state,
and then give it a new occupation state according to
the conditional distribution of \reff{eq.ZRC}
with the other bonds held fixed \cite{note_HB_vs_Metropolis}.
In detail, this means that $e$ will become occupied
with probability $v/(1+v)$ [resp.\ $v/(q+v)$]
in case the endpoints of $e$ are (resp.\ are not)
already connected by a path of occupied bonds not using $e$.
The efficient implementation of this connectivity check
leads to nontrivial algorithmic questions \cite{dynamic_connectivity}
that we will discuss in detail elsewhere \cite{sweeny_fullpaper}.

In two dimensions, the behavior of the ferromagnetic
Potts/random-cluster model is fairly well understood,
thanks to a combination of exact solutions \cite{Baxter_book},
Coulomb-gas methods \cite{Nienhuis_84}
and conformal field theory \cite{DiFrancesco_97}.
But in dimensions $d \ge 3$, many important aspects remain unclear,
including the location of the crossover between
second-order and first-order behavior \cite{fn_1st2nd};
the nature of the critical exponents and their dependence on $q$;
the value of the upper critical dimension for noninteger $q$;
and the qualitative behavior of the critical curve $v_c(q)$ near $q=0$
\cite{forests_3d4d_prl}.
Monte Carlo simulations using the Sweeny \cite{Sweeny_83}
and Chayes--Machta \cite{Chayes-Machta} algorithms
will likely play an important role in elucidating these problems.

In this Letter we present the results of Monte Carlo simulations
using the Sweeny dynamics,
on $d$-dimensional simple hypercubic lattices
of linear size $L$ with periodic boundary conditions.
We shall measure time in units of ``hits'' of a single bond,
but we stress that the natural unit of time is one ``sweep'' of the lattice,
consisting of $dL^d$ hits.
For any observable $\scro$, we define the unnormalized autocorrelation function
at time lag $t$,
\begin{equation}
   C_{\scro} (t)   \;=\;   \< \scro_s  \scro_{s+t} \> - \<\scro\>^2
   \;,
\end{equation}
where expectations are taken in the stationary stochastic process
(i.e., in equilibrium),
and the normalized autocorrelation function
$\rho_{\scro} (t) = C_{\scro} (t) / C_{\scro} (0)$.
We then define the exponential autocorrelation time
\begin{equation}
   \tau_{{\rm exp},\scro} \;=\;
   \limsup_{t \to \pm\infty} {|t| \over   - \log |\rho_{\scro}(t)|}
\end{equation}
and the integrated autocorrelation time
\begin{equation}
   \tau_{{\rm int},\scro}   \;=\;
   \half \sum_{t = -\infty}^{\infty}  \rho_{\scro}(t)
   \;.
\end{equation}
Typically all observables $\scro$
(except those that, for symmetry reasons,
 are ``orthogonal'' to the slowest mode)
have the same value $\tau_{{\rm exp},\scro} = \tau_{{\rm exp}}$.
However, they may have very different amplitudes of ``overlap''
with this slowest mode;
in particular, they may have very different values of the
integrated autocorrelation time,
which controls the efficiency of Monte Carlo simulations
\cite{Sokal_Cargese_96}.
We define dynamic critical exponents $z_{\rm exp}$
and $z_{{\rm int},\scro}$ by
$\tau_{\rm exp} \sim \xi^{z_{\rm exp}}$
and $\tau_{{\rm int},\scro} \sim \xi^{z_{{\rm int},\scro}}$,
where time is measured in ``sweeps''.
On a finite lattice at criticality, $\xi$ can here be replaced by $L$.

In our simulations we measured a variety of observables,
among which are the number of occupied bonds $\scrn = |A|$
and the sum of squares of cluster sizes
$\scrs_2 = \sum |\scrc|^2$.
It is well known \cite{FK_69+72} that $\< \scrs_2 \> = V\chi$,
where $V=L^d$ is the volume and $\chi$ is the Potts-model susceptibility.
A simple variational argument \cite{Li-Sokal}
shows that, in the Sweeny dynamics,
$\tau_{\rm exp} \gtapprox \tau_{{\rm int},\scrn} \ge {\rm const} \times C_H$,
where $C_H$ is the specific heat and time is measured in ``sweeps'';
hence $z_{\rm exp} \ge z_{{\rm int},\scrn} \ge \alpha/\nu$.

\begin{figure}[t]
\vspace*{-1mm}
\begin{center}
\includegraphics[width=\columnwidth]{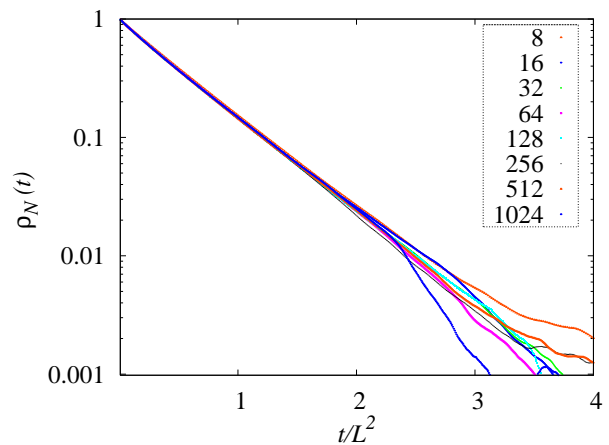}
\end{center}
\vspace*{-6mm}
\caption{
   Autocorrelation function $\rho_{\scrn}(t)$ versus $t/L^2$
   for the critical two-dimensional random-cluster model at $q=0.2$,
   where time $t$ is measured in ``hits''.
}
\label{fig1}
\end{figure}

\begin{figure}[t]
\begin{center}
\includegraphics[width=\columnwidth]{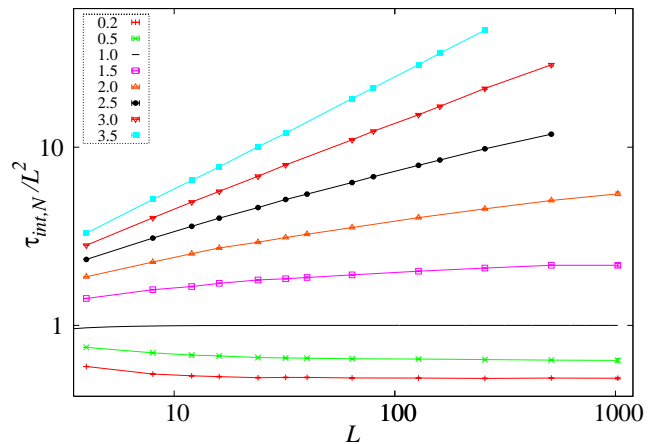}
\end{center}
\vspace*{-6mm}
\caption{
   Integrated autocorrelation times $\tau_{{\rm int},\scrn}$
   for the critical two-dimensional random-cluster model,
   as a function of $q$ and $L$.
   For $q=1$ the analytical result is
   $\tau_{{\rm int},\scrn} = \smhalf (dL^d - 1)$.
}
\label{fig2}
\end{figure}

We began by performing simulations on the square lattice ($d=2$)
at the exact critical point $v_c(q) = \sqrt{q}$ \cite{Baxter_book}
for $q = 0.0005, 0.005, 0.05, 0.2, 0.5, 1, 1.5, 2, 2.5, 3, 3.5$
and a variety of lattice sizes $4 \le L \le 1024$ \cite{note_Metropolis2}.
In all cases the autocorrelation function of $\scrn$
is very close to a pure exponential (Fig.~\ref{fig1}).
The integrated autocorrelation times $\tau_{{\rm int},\scrn}$
are shown as a function of $q$ and $L$ in Fig.~\ref{fig2},
and the corresponding dynamic critical exponents
$z_{\rm exp} \approx z_{{\rm int},\scrn}$ are shown in Table~\ref{table1}.
The estimated exponents are only slightly larger than
the lower bound $\alpha/\nu$, and could conceivably be equal to it
\cite{note_tauintN_over_CH}.
Perhaps surprisingly, these exponents are slightly {\em smaller}\/
than those found recently \cite{cm_2d}
for the Chayes--Machta--Swendsen--Wang \cite{Chayes-Machta} cluster algorithm.

%
%

%
%

%
%

\begin{table}[t]
\begin{center}
\begin{tabular}{|c|cc|cc|c|cc|}
\hline
   $q$  & $z_{{\rm exp}}$  &  $\alpha/\nu$ &
   $w$ &  $r$ &  $z_{{\rm int},\scrs_2}$  &  $d_{\rm red}$ & $d_{\rm F}$  \\
\hline
  0.0005   &    0    &   $-1.9576$ &  0.77  &  4.83  &  $-1.23$  &
      1.2376  &  1.9965 \\
  0.005   &    0    &   $-1.8679$ &  0.79  &  4.18  &  $-1.21$  &
      1.2111  &  1.9891 \\
  0.05   &    0    &   $-1.6005$ &  0.88  &  2.84  &  $-1.12$  &
      1.1299  &  1.9679 \\
  0.2   &    0    &   $-1.2467$ &  0.99  &  1.42  &  $-1.01$  &
      1.0168  &  1.9417 \\
  0.5   &    0    &   $-0.8778$  &  1.11  &  0.80  &  $-0.71$  &
      0.8904  &  1.9172 \\
  1.0   &    0    &   $-0.5000$  &  1.26  &  0.43  &  $-0.32$  &
      0.7500  &  1.8958 \\
  1.5   &    0    &   $-0.2266$  &  1.36  &  0.25  &  $-0.16$  &
      0.6398  &  1.8832 \\
  2.0   & 0 (log)  &    0 (log)  &  1.49  &  0.15  &  $-0.08$  &
      0.5417  &  1.8750 \\
  2.5   & 0.26(1)  &   0.2036  &  1.64  &  0.10  &   0.20  &
      0.4474  &  1.8697 \\
  3.0   & 0.45(1)  &   0.4000  &  1.84  &  0.06  &   0.41  &
      0.3500  &  1.8667 \\
  3.5   & 0.636(2) &   0.6101  &  2.04  &  0.04  &  0.61  &
      0.2375  &  1.8662 \\
\hline
\end{tabular}
\end{center}
\vspace*{-3mm}
\caption{
   Estimated dynamic critical exponents
   for the two-dimensional random-cluster model as a function of $q$.
   Specific-heat exponent $\alpha/\nu$ and
   red-bond (resp.\ whole-cluster) fractal dimension
   $d_{\rm red}$ (resp.\ $d_{\rm F}$) are shown for comparison
   \cite{note_red_bonds}.
}
\label{table1}
\end{table}

\begin{figure}[t]
\vspace*{-3mm}
\begin{center}
\includegraphics[width=\columnwidth]{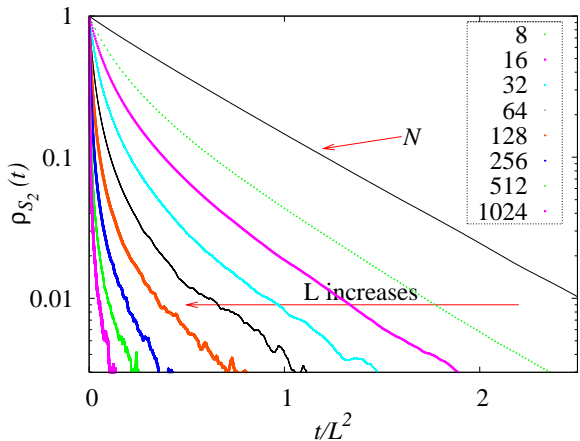}
\end{center}
\vspace*{-6mm}
\caption{
   Autocorrelation function $\rho_{\scrs_2}(t)$ versus $t/L^2$
   for the critical two-dimensional random-cluster model at $q=0.2$.
   The autocorrelation function $\rho_{\scrn}(t)$ is shown for comparison.
}
\label{fig3}
\end{figure}
\begin{figure}[t]
\vspace*{-2mm}
\begin{center}
\includegraphics[width=\columnwidth]{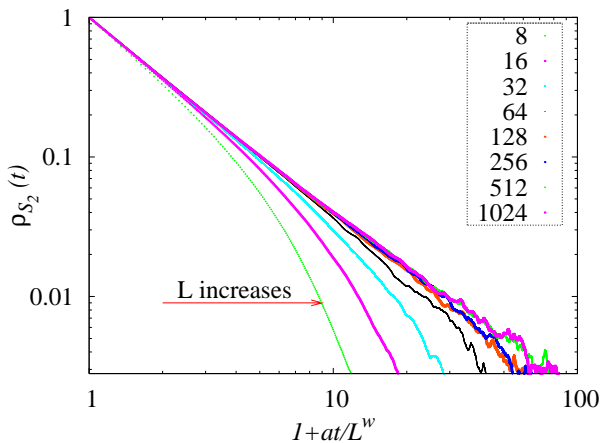}
\end{center}
\vspace*{-6mm}
\caption{
   Autocorrelation function $\rho_{\scrs_2}(t)$ versus $1 + at/L^w$
   for the critical two-dimensional random-cluster model at $q=0.2$,
   with $w = 0.99$ and $a = 0.55$.
}
\label{fig4}
\end{figure}

A more interesting and unusual dynamic behavior
is exhibited by the observable $\scrs_2$.
In Fig.~\ref{fig3} we plot
the autocorrelation function $\rho_{\scrs_2}(t)$ for $q=0.2$.
Clearly $\rho_{\scrs_2}(t)$ exhibits a fast decay
in a time much {\em less}\/ than a single sweep
(i.e.\ of order $L^w$ for some $w < 2$)
as a prelude to the ultimate exponential decay $e^{-t/\tau_{\rm exp}}$.
To analyze this short-time behavior,
we plot $\rho_{\scrs_2}(t)$ versus $t/L^w$
and adjust the exponent $w$ until all the points fall on a scaling curve
$\rho_{\scrs_2}(t) = f(t/L^w)$ in the limit $L \to\infty$.
We find $w \approx 0.99$.
Furthermore, the function $f$ is very close to $f(x) = (1+ax)^{-r}$
with $a = 0.55$ and $r = 1.42$ (Fig.~\ref{fig4});
in any case it behaves like $f(x) \sim x^{-r}$ as $x \to \infty$.

%
%

%
%

\begin{figure}[t]
\vspace*{-3mm}
\begin{center}
\includegraphics[width=\columnwidth]{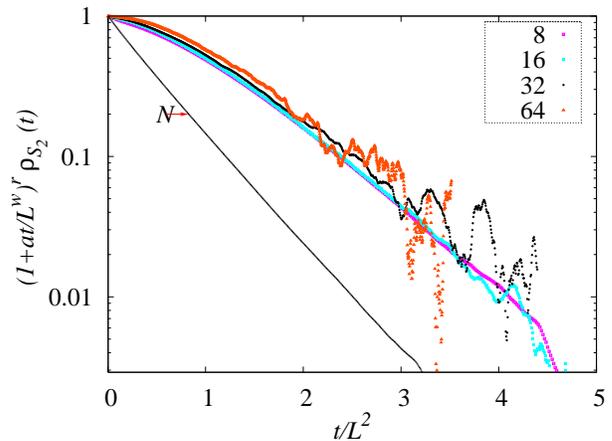}
\end{center}
\vspace*{-6mm}
\caption{
   $(1+t/L^w)^r \rho_{\scrs_2}(t)$ versus $t/L^2$
   for the critical two-dimensional random-cluster model at $q=0.2$,
   with $w = 0.99$, $a = 0.55$ and $r=1.42$.
   The autocorrelation function $\rho_{\scrn}(t)$ is shown for comparison.
}
\label{fig5}
\end{figure}

Finally we can analyze the universal crossover from
short-time to long-time behavior, which we hypothesize is of the form
$\rho_{\scrs_2}(t) = f(t/L^w) g(t/L^{d+z_{\rm exp}})$,
by plotting $(1+at/L^w)^r \rho_{\scrs_2}(t)$ versus $t/L^{d+z_{\rm exp}}$.
A fairly clear scaling curve is seen (Fig.~\ref{fig5}),
though it is noisy for large lattices.
Using this scaling Ansatz to compute the area under the curve of
$\rho_{\scrs_2}(t)$, we conclude that
\begin{equation}
   z_{{\rm int}, \scrs_2} \;=\;
      \cases{
          r(w-d) + (1-r)z_{\rm exp}   &  if $r < 1$  \cr
          w-d                         &  if $r > 1$
      }
\end{equation}

%
%

Similar analyses for the other values of $q$
yield the exponents reported in Table~\ref{table1}.
Note that critical speeding-up ($w < d$)
and critical slowing-down ($z_{\rm exp} > 0$) can coexist.

Critical speeding-up also occurs in the random-cluster model
in dimensions $d > 2$.
We simulated the three-dimensional random-cluster model \cite{note_Metropolis2}
at the estimated critical points
$w = v/q = 0.43365$ for $q=0$ \cite{forests_3d4d_prl},
$p = v/(1+v) = 0.2488126$ for $q=1$ \cite{d=3_percolation},
and $v = e^{2\beta} - 1$ with $\beta = 0.22165455$ for $q=2$ \cite{Deng_03},
using lattice sizes $4 \le L \le 256$ for $q=0,1$
and $4 \le L \le 64$ for $q=2$.
The qualitative behavior was the same as in $d=2$,
and the estimated exponents are shown in Table~\ref{table2}.
Our value of $z_{\rm exp}$ for $q=2$ is consistent with that of
Wang, Kozan and Swendsen \cite{Sweeny_83}.
In fact, $\tau_{{\rm int},\scrn}/C_H$ for $q=2$ is close to constant,
so it is conceivable that $z_{\rm exp} = \alpha/\nu$ exactly.

%
%

\begin{table}[t]
\begin{center}
\begin{tabular}{|c|cc|cc|c|cc|}
\hline
   $q$  & $z_{{\rm exp}}$  &  $\alpha/\nu$ &
   $w$ &  $r$ &  $z_{{\rm int},\scrs_2}$  &  $d_{\rm red}$ & $d_{\rm F}$  \\
\hline
  0   &    0    &   $-1.44(5)$ &  1.52  &  1.04  &  $-1.48$  &
      ?       &  2.5838(5) \\
  1   &    0    &   $-0.713(1)$ &  1.87  &  0.32  &  $-0.36$  &
      1.1437(6)  &  2.5219(2) \\
  2   &  0.35(1)  &   $0.174(2)$ &  2.55  &  0.08  &  0.29  &
      0.757(2)  &  2.4816(1) \\
\hline
\end{tabular}
\end{center}
\vspace*{-3mm}
\caption{
     Same as Table~\ref{table1},
     for the three-dimensional random-cluster model.
     $\alpha/\nu$ and $d_{\rm F}$ from \cite{forests_3d4d_prl} \cite{Deng_05}
          \cite{Pelissetto-Vicari,Deng_03} for $q=0,1,2$;
     $d_{\rm red}$ from \cite{Deng_05} \cite{Deng_04} for $q=1,2$.
}
\label{table2}
\end{table}


In retrospect it is not surprising that a ``global'' observable like $\scrs_2$
could exhibit significant decorrelation in a time much less than one sweep.
After all, FK clusters are fractals:
a large cluster can sometimes be broken into two large pieces
by one or a few bond deletions,
and two large clusters can sometimes be joined
by one or a few bond insertions.
This reasoning suggests correctly that the critical speeding-up
should be strongest when the cluster is most fragile,
i.e.\ the red-bond fractal dimension $d_{\rm red}$ \cite{note_red_bonds}
is largest, namely at small $q$.

We can pursue this idea further and suggest that
the decorrelation of $\scrs_2$ is caused principally
by hitting a few (order~1) red bonds:
this takes a time $\propto L^{d-d_{\rm red}}$,
so we predict $w = d-d_{\rm red}$.
Our data (Tables~\ref{table1} and \ref{table2})
are in amazingly good agreement with this prediction
for $q \ltapprox 2$, i.e.\ when $z_{\rm exp} = 0$.
However, they deviate from it when $z_{\rm exp} > 0$,
for reasons that we do not yet understand.
Note in particular that for $d \ge 6$
and $q = 1$ (resp.\ $0 \le q \le 1$),
we expect $d_{\rm red} = y_{t0} = 2$ (resp.\ $d_{\rm red} \ge 2$)
and hence $w = d-2$ (resp.\ $w \le d-2$).

We lack, at present, any theory (or even any numerology) for $r$.
But in two dimensions it seems that $r \to 5$ (resp.\ 0)
as $q \to 0$ (resp.\ 4).

Let us note that a similar ``two-time-scale'' behavior
is observed in the pivot dynamics \cite{Madras_88}
for ordinary random walk (or self-avoiding walk),
in which ``global'' observables such as the end-to-end distance
and the radius of gyration exhibit a fast relaxation
$\tau_{\rm short} \sim N^0$ while the slowest mode has
$\tau_{\rm exp} \sim N$ (here $N$ is the number of steps in the walk).
Indeed, it is conceivable that {\em most}\/ types of dynamics ---
perhaps even single-spin-flip (Glauber) dynamics ---
exhibit this two-time-scale effect
(i.e., $w < d+z_{\rm exp}$ and hence $z_{{\rm int},\scro} < z_{\rm exp}$)
to a greater or lesser extent.

On a practical level, our results show that the Sweeny algorithm is,
despite its local nature,
an unexpectedly efficient algorithm for simulating the random-cluster model.
Of course, for $0 < q < 1$ it is the {\em only}\/ known algorithm.
For $q \ltapprox 1.5$ its efficiency is enhanced
by strong critical speeding-up.
Even for larger values of $q$, it is a potential competitor to the
Chayes--Machta \cite{Chayes-Machta} cluster algorithm
if efficient connectivity-checking algorithms can be found
\cite{dynamic_connectivity,sweeny_fullpaper}.

Details of these simulations and their data analysis
will be reported separately \cite{sweeny_fullpaper}.

\begin{acknowledgments}
This work was supported in part by NSF grants PHY--0116590 and PHY--0424082.
\end{acknowledgments}



\begin{thebibliography}{99}
\bibitem{note_HB_vs_Metropolis}  This is the ``heat-bath'' version
of the Sweeny dynamics, in which the new occupation state of the bond $e$
is independent of the old occupation state.
A slightly more efficient alternative is the ``Metropolis'' version,
in which a proposal to flip the occupation state of $e$
is accepted or rejected according to the Metropolis criterion.
\bibitem{note_Metropolis2}  We used the ``Metropolis'' version
   of the Sweeny algorithm:  see \cite{note_HB_vs_Metropolis} above.
\end{thebibliography}
\end{document}